\documentclass[a4paper,10pt,twoside,bibnote]{cpc-hepnp}

\usepackage{multicol}
\usepackage{float}

\usepackage{booktabs}
\usepackage{footnote}

\usepackage{lipsum}
\usepackage{amsmath}    
\usepackage{amssymb}    
\usepackage{color}         
\usepackage{graphicx}     
\usepackage[T1]{fontenc}
\usepackage{times}         
\usepackage{mathrsfs} 
\usepackage{myterms}
\usepackage{hyperref} 
\hypersetup{
colorlinks=true,       
linkcolor=blue,        
citecolor=blue,        
filecolor=magenta,     
urlcolor=blue          
}
\usepackage[all]{hypcap} 

\usepackage[none]{hyphenat} 

\usepackage{color,xcolor,fancybox,epsf,rotating,colordvi}

\usepackage{ulem}

\usepackage[all]{hypcap} 

\begin{document}


\title{Light Higgs boson in the NMSSM confronted with the CMS\\ diphoton and ditau excesses\thanks{supported by the National Natural Science Foundation of China (NNSFC) under grant Nos. 12275066, 11605123, 11547103, and 12074295.}}

\author{
			Weichao Li$^{1}$%
			\quad Haoxue Qiao$^{1;1)}$
			\quad Jingya Zhu$^{2;2)}$
			\email{qhx@whu.edu.cn (corresponding author)}%
			\email{zhujy@henu.edu.cn (corresponding author)}%
		}

\maketitle

\address{
			$^1$ School of Physics and Technology, Wuhan University, Wuhan 430072, China \\
			$^2$ Joint Center for Theoretical Physics, and School of Physics and Electronics, Henan University, Kaifeng 475004, China \\
		}
		

\begin{abstract}
In 2018, the CMS collaboration reported a di-photon excess around 95.3 GeV with a local significance of 2.8 $\sigma$. 
Interestingly, the CMS collaboration also reported a di-tau excess recently at $95\sim 100 \GeV$ with a local significance of $2.6\sim 3.1 ~\sigma$. 
Besides, a $b\bar{b}$ excess at 98 GeV with a 2.3 $\sigma$ local significance was reported with LEP data about twenty years ago. 
In this work, we consider interpreting these excesses together with a light Higgs boson in the next-to-minimal supersymmetric standard model (NMSSM). 
We conclude that in NMSSM the $95\sim 100\GeV$ excesses are difficult to be satisfied simultaneously (not possible globally at $1\sigma$ level, or simultaneously at $2\sigma$ level), and we analyze two partial-satisfied scenarios: the globally $2\sigma$ scenario and small di-photon scenario. 
An approximate equation of global fit to the three excesses is derived, and two representative types of surviving samples are analyzed in detail. 
Since the mass regions of these excesses are near the Z boson, we also consider checking the light Higgs boson in the $t\bar{t}$-associated channels. 
The detailed results may be useful for further checking the low-mass-region excesses in the future. 
\end{abstract}

		\begin{keyword}
			Higgs boson, supersymmetry phenomenology, NMSSM
		\end{keyword}
		
		\begin{pacs}
			\qquad     {\bf DOI:} 
		\end{pacs}

\begin{multicols}{2}

\section{Introduction}
\label{sec:intro}

In 2012, the ATLAS and CMS collaborations reported a new boson of around 125 GeV discovered at the LHC \cite{ATLAS:2012yve, CMS:2012qbp}. 
Then it was proved to be the Standard Model (SM)-like Higgs boson, according to its spin, $CP$ property, production and decay performances in Run I and Run II data globally \cite{ATLAS:2022vkf, CMS:2022dwd, ATLAS:2016neq}. 
Higgs boson is related to the electroweak symmetry-breaking mechanism, hierarchy problem, and interesting phenomenology in many new physics models.  
Whether there are additional Higgs bosons is a natural, important, and unsolved question. 
So ten years after the 125-GeV Higgs boson was discovered, experimentalists are still making efforts to search for additional Higgs scalars, even if in the low-mass region. 

In 2018, the CMS collaboration reported a di-photon excess around 95.3 GeV with a local significance of 2.8 $\sigma$  \cite{CMS:2018cyk}, with a signal strength of 
\begin{eqnarray}
R_{\gamma\gamma}^{\rm ex} = \frac{\sigma^{\rm ex}(gg \to \phi \to \gamma\gamma)}{\sigma^{\rm SM}(gg \to h \to \gamma\gamma)} = 0.6 \pm 0.2 \;.
\end{eqnarray}
Interestingly, the CMS collaboration also reported a di-tau excess recently at $95\sim 100 \GeV$ with a local significance of $2.6\sim 3.1 ~\sigma$ \cite{CMS:2022goy}, with the signal strength of  
\begin{eqnarray}
R_{\tau\tau}^{\rm ex} = \frac{\sigma^{\rm ex}(gg \to \phi \to \tau^+\tau^-)}{\sigma^{\rm SM}(gg \to h \to \tau^+\tau^-)} = 1.2 \pm 0.5 \;.
\end{eqnarray}
Besides, a $b\bar{b}$ excess around 98 GeV with a local significance of 2.3 $\sigma$ was reported with the LEP data about twenty years ago \cite{LEPWorkingGroupforHiggsbosonsearches:2003ing}, 
whose signal strength is  
\begin{eqnarray}
R_{bb}^{\rm ex} = \frac{\sigma^{\rm ex}(e^+e^- \to Z\phi \to Zb\overline b)}{\sigma^{\rm SM}(e^+e^- \to Zh \to Zb\overline b)} = 0.117 \pm 0.057 \;.
\end{eqnarray}

Since the three excesses are near in mass regions, and are comparable in signal strengths with a SM Higgs boson of the same mass, a series of works attempt to interpret them by one additional Higgs-like scalar in new physics models, with \cite{Biekotter:2022jyr, Biekotter:2022abc, Iguro:2022dok, Iguro:2022fel} and without \cite{Cline:2015lqp, Cao:2016uwt, Fox:2017uwr, Biekotter:2017xmf, Haisch:2017gql,  Liu:2018xsw, Liu:2018ryo, Hollik:2018yek, Domingo:2018uim,  Biekotter:2019kde, Cao:2019ofo, Cline:2019okt, Sachdeva:2019hvk,  Biekotter:2020cjs, Aguilar-Saavedra:2020wrj, Biekotter:2021ovi, Heinemeyer:2021msz, Biekotter:2021qbc, Benbrik:2022azi, Benbrik:2022dja} the di-tau excess. 

Supersymmetry (SUSY) \cite{Fayet:1976et, Fayet:1977yc, Martin:1997ns} is a popular theory beyond the SM. 
The next-to-minimal supersymmetric standard model (NMSSM) \cite{Ellwanger:2009dp} includes two Higgs doublets and one singlet, thus having more freedom than the minimal supersymmetric standard model (MSSM) in the Higgs sector \cite{Maniatis:2009re}. 
It can naturally accommodate a SM-like Higgs boson of 125 GeV with signal strengths fitting the experimental data well \cite{Cao:2012fz, Ellwanger:2011aa, King:2012is, Kang:2012sy, Gunion:2012zd, Benbrik:2012rm, King:2012tr, Ellwanger:2012ke, Cao:2012yn, Wang:2020tap}, and can in addition predict a kind of Higgs exotic decay to a pair of additional Higgs scalar lighter than the half mass \cite{Cao:2013gba, Huang:2013ima, Curtin:2013fra, Ma:2020mjz}. 
Since now we have three possible excesses in different channels in the $95\sim 100 \GeV$ region, it is interesting to see whether it is possible to interpret all three excesses together in the NMSSM. 
In this work, we impose the three excesses from one $95\sim 100 \GeV$ Higgs scalar in NMSSM, investigating its status confronting the excesses. 
In our calculations, we also consider other related constraints including Higgs data, SUSY searches, dark matter relic density and direct detection, etc.

The rest of this paper is organized as follows.
In Sec. \ref{sec:ana} we introduce the Higgs sector in NMSSM and present the relevant analytic equations briefly.
In Sec. \ref{sec:num} we show the numerical-calculation results and discussions.
Finally, we draw our main conclusions in Sec. \ref{sec:conc}.

\section{The Higgs sector in NMSSM}
\label{sec:ana}
SUSY models are mainly determined by their superpotential and soft-breaking terms. 
In the NMSSM, they can be written as 
\begin{eqnarray}
W \!\!\!\!\!&=&\!\!\!\!\! W_{{\rm MSSM}}^{\mu\to \lambda \hat{S}} 
+\kappa \hat{S}^3 /3,\\
V_{\rm soft} \!\!\!\!\!&=&\!\!\!\!\! \tilde m_{H_u}^2|H_u|^2 + \tilde m_{H_d}^2|H_d|^2
+\tilde m_S^2|S|^2 
\nonumber\\
&& \!\!\!\!\! +( \lambda A_{\lambda} SH_u\cdot H_d
+\kappa A_{\kappa} S^3 /3 + h.c.) \,,
\end{eqnarray}
where $W_{{\rm MSSM}}^{\mu\to \lambda \hat{S}}$ is the MSSM superpotential with the $\mu$-term generated effectively by the Vacuum Expectation Value (VEV) of singlet field, and $\tilde{m}_{H_u}$, $\tilde{m}_{H_d}$, $\tilde{m}_{S}$, $A_\lambda$ and $A_\kappa$ are soft-breaking parameters. 
$\hat{H_u}$, $\hat{H_d}$ are the SU(2) doublet and $\hat{S}$ is the singlet Higgs superfields, and after getting VEVs the scalar fields can be expressed as
\begin{eqnarray}
&&H_{u}=\left(
\begin{array}{c}
H^{+}_{\rm u} \\
v_{u}+ \frac{\phi_{u}+i\varphi_{u}}{\sqrt{2}} \\
\end{array} 
\right) ,
\qquad
H_{d}=\left(
\begin{array}{c}
v_{d}+ \frac{\phi_{d}+i\varphi_{d}}{\sqrt{2}} \\
H^{-}_{\rm d} \\
\end{array}
\right)   ,
\nonumber\\
&&S=v_{s}+ \frac{\phi_{s}+i\varphi_{s}}{\sqrt{2}} \,, \label{H-F}
\end{eqnarray}
then the parameter $\tan\beta \equiv v_u/v_d$. 

The three gauge-eigenstate scalars $\{\phi_u, \phi_d, \phi_s\}$ mix to form three CP-even mass-eigenstate Higgs scalars $\{h_1, h_2, h_3\}$, with mass order $m_{h_1}<m_{h_2}<m_{h_3}$, and the mixing matrix $\{S_{ij}\}_{3\times 3}$: 
\begin{equation}
\left(
\begin{array}{ccc}
h_1\\
h_2\\
h_3\\
\end{array}
\right)
=
\left(
\begin{array}{ccc}
S_{11} & S_{12} & S_{13} \\
S_{21} & S_{22} & S_{23} \\
S_{31} & S_{32} & S_{33} \\
\end{array}
\right)
\left(
\begin{array}{ccc}
\phi_u\\
\phi_d\\
\phi_s\\
\end{array}
\right)
\end{equation}
The reduced couplings of $h_1$ to up- and down-type fermions, and massive gauge bosons, are given by
\begin{eqnarray}
&&c_t=S_{11}/\sin\beta \,, \nonumber\\
&&c_b=S_{12}/\cos\beta \,, \nonumber\\
&&c_V=S_{11}\sin\beta +S_{12}\cos\beta \,.
\label{h1coup}
\end{eqnarray}
While the loop-induced coupling to gluons $c_g$ is mainly determined by $c_t$ and light colored SUSY particles, and that of photon $c_\gamma$ are mainly by $c_t$, $c_V$, and light charged SUSY particles.

\section{Numerical results and discussions}
\label{sec:num}

In the calculation, we first perform a scan over the parameter space of NMSSM with the public code $\textsf{NMSSMTools\_5.6.1}$ \cite{Ellwanger:2004xm, Ellwanger:2005dv, Ellwanger:2006rn} under a series of experimental and theoretical constraints\footnote{We do not take the constraints to Higgs couplings into account, for these constraints are only global fit results under some assumptions, e.g., no exotic and invisible Higgs decays when varying the couplings. Instead, we use direct experimental constraints to Higgs signals with the code {\sf HiggsSignals}.}. 
The parameter space we consider are: 
\begin{eqnarray}\label{NMSSM-scan}
	&& 0.1 \textless \lambda\textless 0.7,
	~~~|\kappa| \textless 0.7,~~~
	1\textless\tan\beta \textless 60,
	\nonumber\\
	&&M_{0},|M_{3}|,|A_{0}|,|A_{\lambda}|,|A_{\kappa}|\textless 10 {\rm ~TeV} \,,\nonumber\\
	&& \mu_{\rm eff}, |M_{1}|, |M_{2}|\textless 1 {\rm ~TeV} \,.
\end{eqnarray}
Note that the NMSSM we consider in this work is GUT-scale constrained, where both Higgs and gaugino masses are considered non-universal. So $M_0$ and $A_0$ are the unified sfermion masses and trilinear couplings in the sfermion sector, and $M_{1,2,3}$ are the gaugino masses at the GUT scale. 
While the three non-universal Higgs masses at the GUT scale are calculated from the minimization equations, with $\lambda$, $\kappa$ and $\mu_{\rm eff}\equiv \lambda v_S$ at the SUSY scale as the input parameters.
The $\mu_{\rm eff}$ parameter is chosen to be positive to interpret the muon $g-2$ anomaly. 
One sign in three of $M_{1,2,3}$ can be absorbed in a field redefinition \cite{Martin:1997ns}. The sign of $M_3$ can have other effects, e.g., as shown in Ref.\cite{Wang:2018vrr}. 

The constraints we imposed include: 
(i) A SM-like Higgs boson\footnote{We calculate Higgs masses and their mixing with 
the most complete calculation implemented in {\sf NMSSMTools}, which includes full one-loop and dominant two-loop corrections.
}
with mass around 125 GeV (i.e., 123 $\sim$ 127 GeV) and signal strengths accord with the latest data in $\textsf{HiggsSignals-2.2.3beta}$ \cite{Bechtle:2020uwn, Bechtle:2013xfa}. 
(ii) The exclusion limits in searching for additional Higgs boson at the LEP, Tevatron, and LHC, which are collected in  $\textsf{HiggsBounds-5.10.1}$ \cite{Bechtle:2020pkv, Bechtle:2013wla, Bechtle:2008jh}. 
(iii) The upper limit of dark matter relic density with uncertainty ($\Omega h^2 \leq 0.131$) \cite{Tanabashi:2018oca, Hinshaw:2012aka, Ade:2013zuv}, and direct detections \cite{Aprile:2018dbl}, where the quantities are calculated with {\sf micrOMEGAs} inside {\sf NMSSMTools}.
(iv) The exclusion limits in SUSY searches imposed in {\sf SModelS-v2.1.1} \cite{Kraml:2013mwa, Ambrogi:2018ujg, Khosa:2020zar, Alguero:2021dig}, such as electroweakinos in multilepton channels \cite{CMS:2017moi, CMS:2018szt}, gluino and first-two-generation squarks  \cite{ParticleDataGroup:2022pth}, etc. 
(v) Theoretical constraints of vacuum stability and no Landau pole below GUT scale \cite{Ellwanger:2006rn}.

To interpret the CMS di-photon and di-tau, and LEP $b\bar{b}$  excesses together, we also require a light Higgs boson of $95\sim100\GeV$. 
For the surviving samples, we define a chi-square quantity $\chi^2_{\gamma\gamma+\tau\tau+bb}$ to describe its ability to interpret the three excesses globally. 
\begin{eqnarray}
	\chi^2_{\gamma\gamma+\tau\tau+bb} = \chi_{\gamma\gamma}^2 + \chi_{\tau\tau}^2 + \chi_{bb}^2 \, ,
\end{eqnarray}
where
\begin{eqnarray}
	\chi_{i}^2 = \left(
	\frac{R_{i}-\bar{R}_{i}^{\rm ex}}
	{\delta\! R_{i}^{\rm ex}}
	\right)^2
	\,, \nonumber
\end{eqnarray}
with $i=\gamma\gamma,\tau\tau,bb$, $R_{i}$ standing for the corresponding theoretical signal strength of our samples, $\bar{R}_{i}^{\rm ex}$ and $\delta\! R_{i}^{\rm ex}$ for the corresponding experimental mean and error values, respectively. 
With $\chi^2_{\gamma\gamma+\tau\tau+bb}\le 8.03$, the surviving samples can interpret the three excesses globally at $2\sigma$ level, and are called `globally $2\sigma$ samples', or simplified as `$2\sigma$ samples' hereafter. 
We note that for surviving samples the minimum value of $\chi^2_{\gamma\gamma+\tau\tau+bb}$ is $5.37$, so there are no samples satisfying the three excesses at $1\sigma$ level globally ($\chi^2_{\gamma\gamma+\tau\tau+bb} < 3.53$).
In Tab. \ref{table1} we list the parameter regions for the  $2\sigma$ and all surviving samples respectively. 

\begin{table}[H]
\centering
\tabcaption{\label{table1}The parameter regions for the $2\sigma$ ($\chi^2_{\gamma\gamma+\tau\tau+bb}\le 8.03$) and all surviving samples respectively.}
	\begin{tabular}{ccc}
		\toprule
		\hline
		~~~~~~~~~~~~&~~~ ~~~$2\sigma$ samples~~~ ~~~& ~~~all surviving samples ~~~\\
		\hline
		$\lambda$           &       0.11 $\sim$ 0.58       &  0.10 $\sim$ 0.69\\
		$\kappa$            &       -0.60 $\sim$ 0.55      &  -0.56 $\sim$ 0.61\\
		tan$\beta$          &       6.4 $\sim$ 45.2          &  2.6 $\sim$ 50.6\\
		$\mu_{\rm eff}$/GeV &       139 $\sim$ 487       &  102 $\sim$ 978\\
		$M_0$/TeV           &       0 $\sim$ 9.5       &  0 $\sim$ 10.0\\
		$A_0$/TeV           &       -5.0 $\sim$ 6.5    &  -8.3 $\sim$ 9.3\\
		$M_1$/GeV           &       -805 $\sim$ 199     &  -1000 $\sim$ 993\\
		$M_2$/TeV           &       -6.7 $\sim$ 1.0     &  -10.0 $\sim$ 2.4\\
		$M_3$/TeV           &       -3.4 $\sim$ 6.4    &  -4.8 $\sim$ 9.8\\
		$A_\lambda$/TeV     &       1.4 $\sim$ 10.0    &  0.1 $\sim$ 10.0\\
		$A_\kappa$/TeV      &       -2.0 $\sim$ 2.4    &  -2.7 $\sim$ 2.7\\
		\hline
		\bottomrule
	\end{tabular}
\end{table}	

\begin{figure*}[!htbp]
\centering
\includegraphics[width=0.96\textwidth]{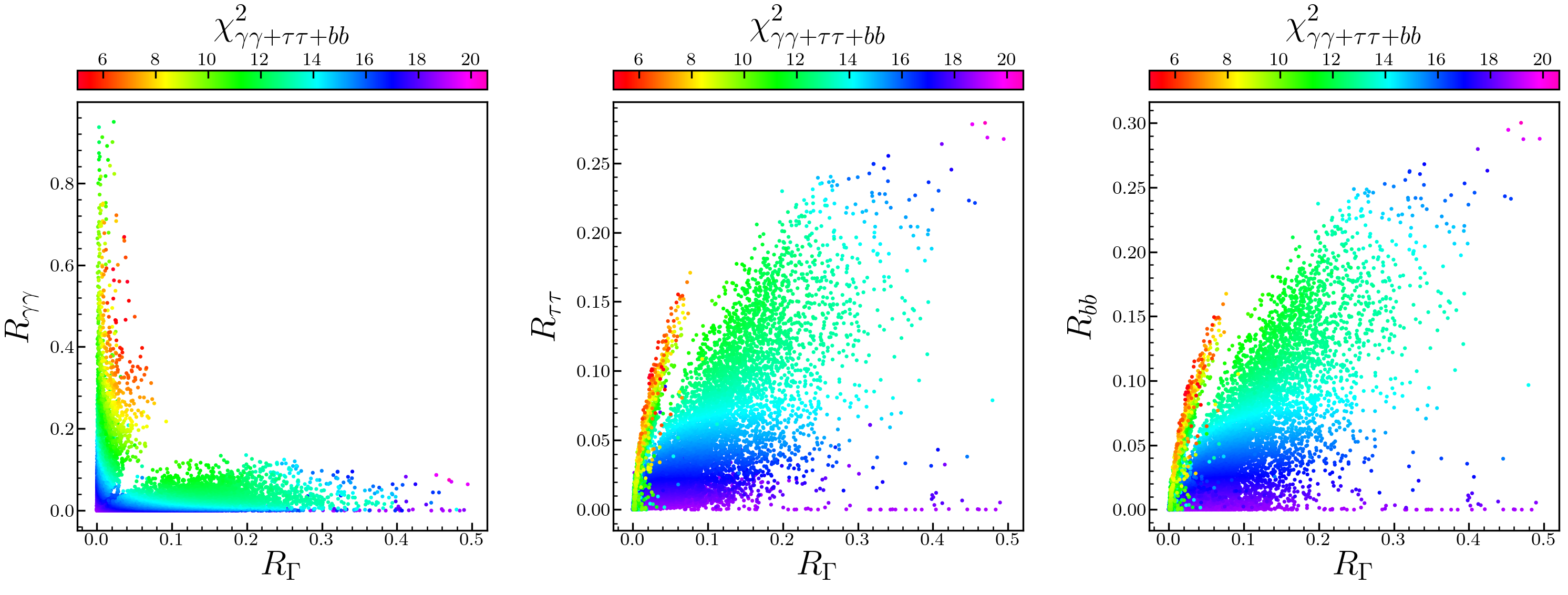}
\caption{Surviving samples on the planes of signal strength $R_{\gamma \gamma}$ ($gg \to h_1 \to \gamma\gamma$) (left), $R_{\tau\tau}$ ($gg \to h_1\to \tau \bar{\tau}$) (middle), $R_{bb}$($e^{+}e^{-}\to Zh_1\to Zb\bar{b}$) (right) versus width ratio $R_{\Gamma}$, respectively, with colours indicate $\chi^2_{\gamma\gamma+\tau\tau+bb}$. }
\label{fig1}
\end{figure*}

\begin{figure*}[!htbp]
\centering
\includegraphics[width=0.96\textwidth]{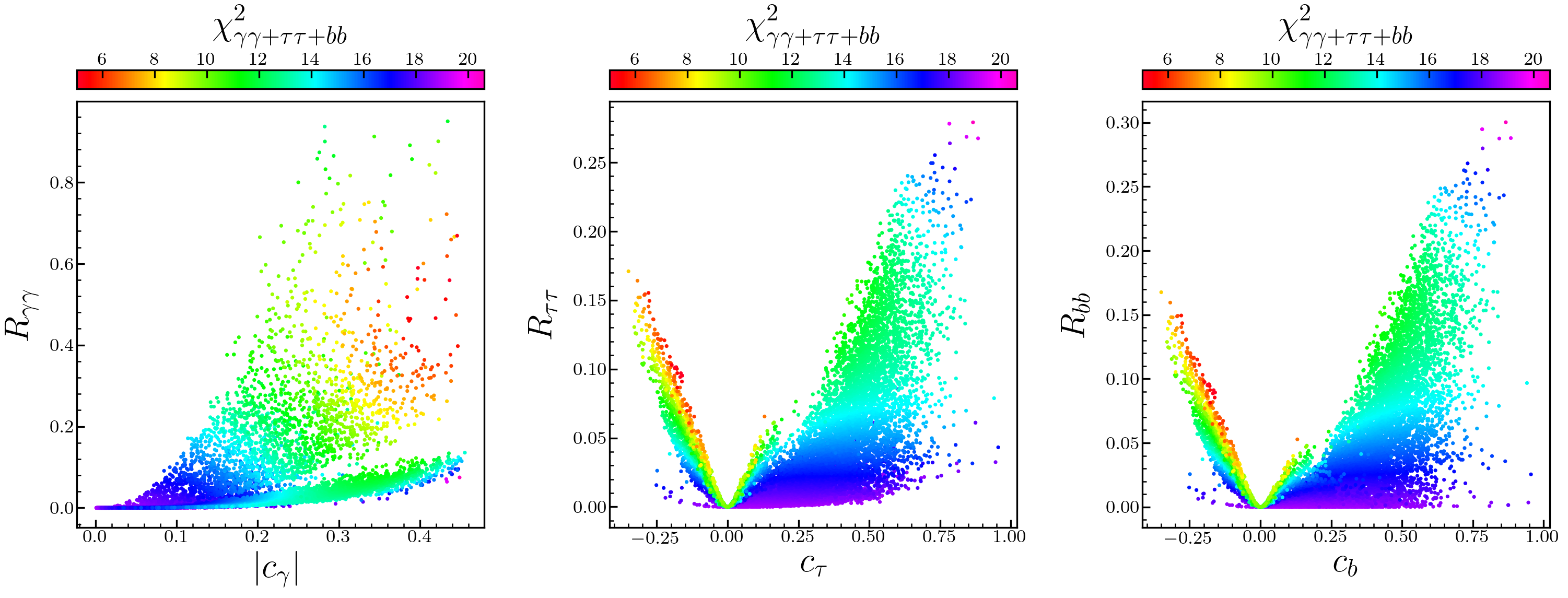}\\
\includegraphics[width=0.96\textwidth]{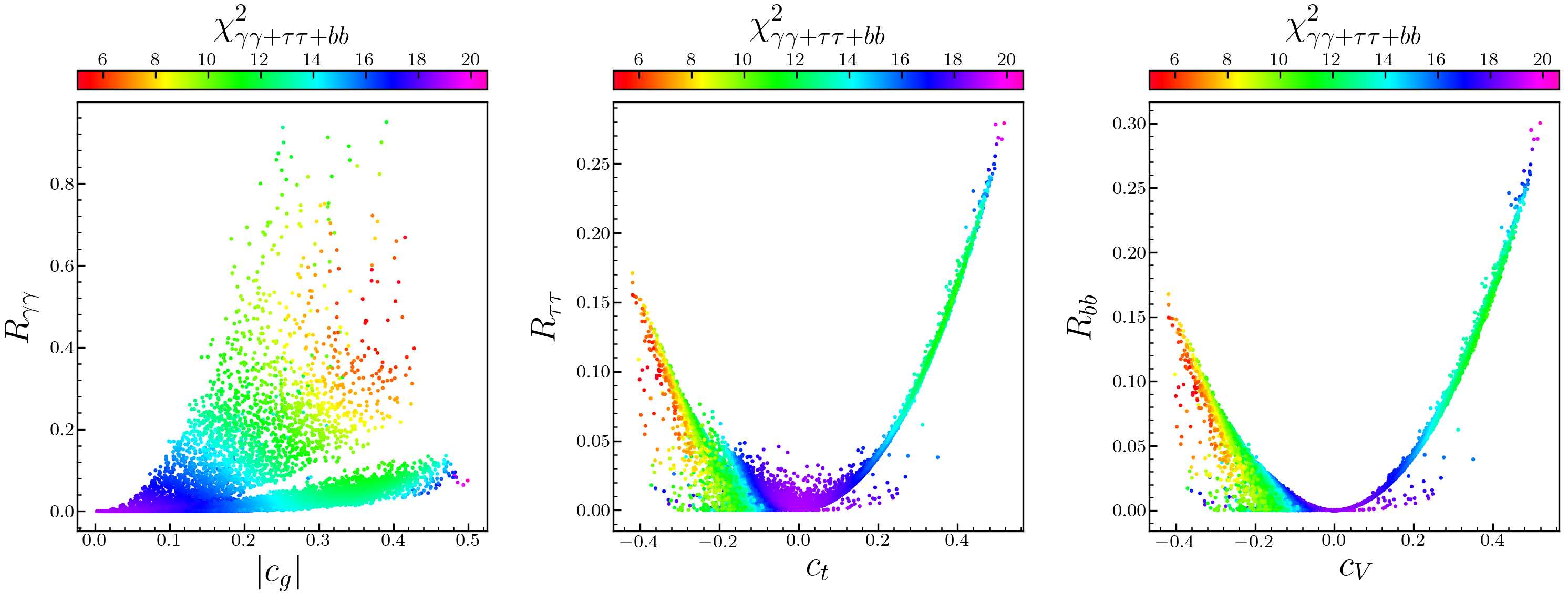}
\caption{Same as in Fig. \ref{fig1}, but on the planes of signal strength versus reduced coupling: 
$R_{\gamma\gamma}$ versus $c_{\gamma}$ (upper left), $R_{\gamma\gamma}$ versus $c_{g}$ (lower left), $R_{\tau\tau}$ versus $c_{\tau}$ (upper middle), $R_{\tau\tau}$ versus $c_{t}$ (lower middle), $R_{bb}$ versus $c_{b}$ (upper right), $R_{bb}$ versus $c_{V}$ (lower right).}
\label{fig2}
\end{figure*}

\begin{figure*}[htb]
\centering
\includegraphics[width=0.96\textwidth]{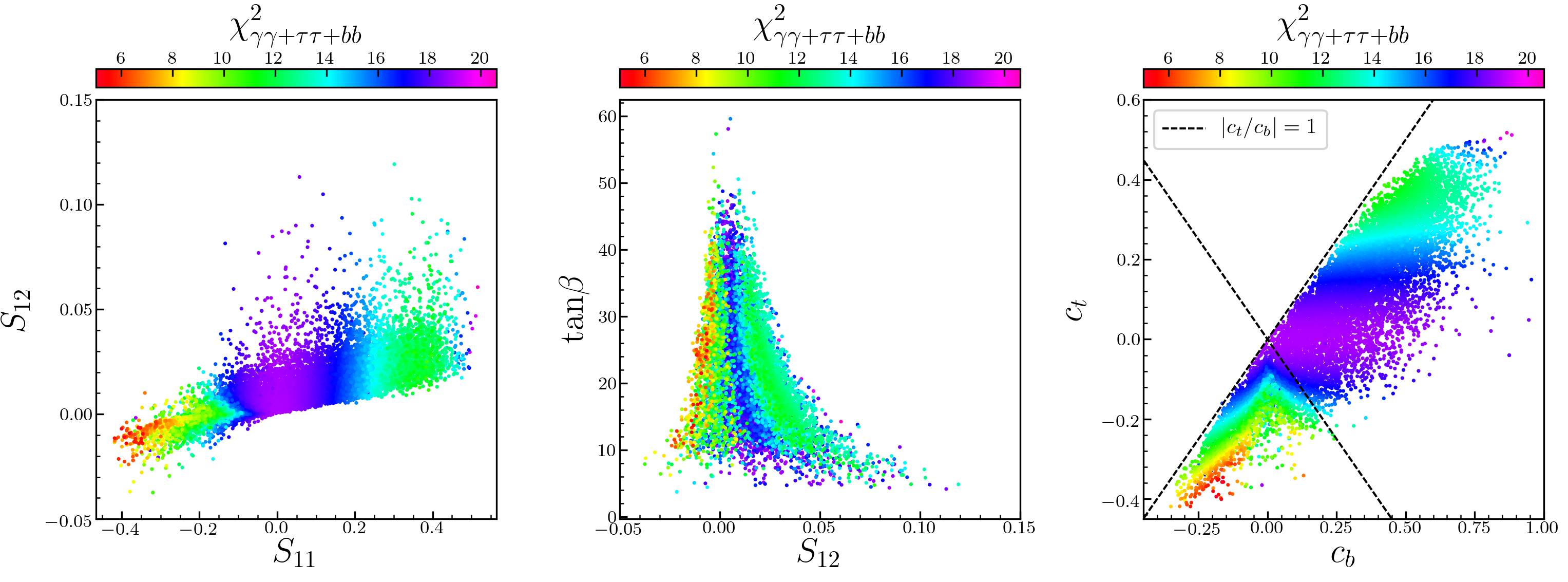}
\caption{Same as in Fig. \ref{fig1}, but on $S_{12}$ versus $S_{11}$ (left), $\tan\beta$ versus $S_{12}$ (middle), and $c_t$ versus $c_b$ (right) planes, respectively.}
\label{fig3}
\end{figure*}

\begin{figure*}[htb]
\centering
\includegraphics[width=0.96\textwidth]{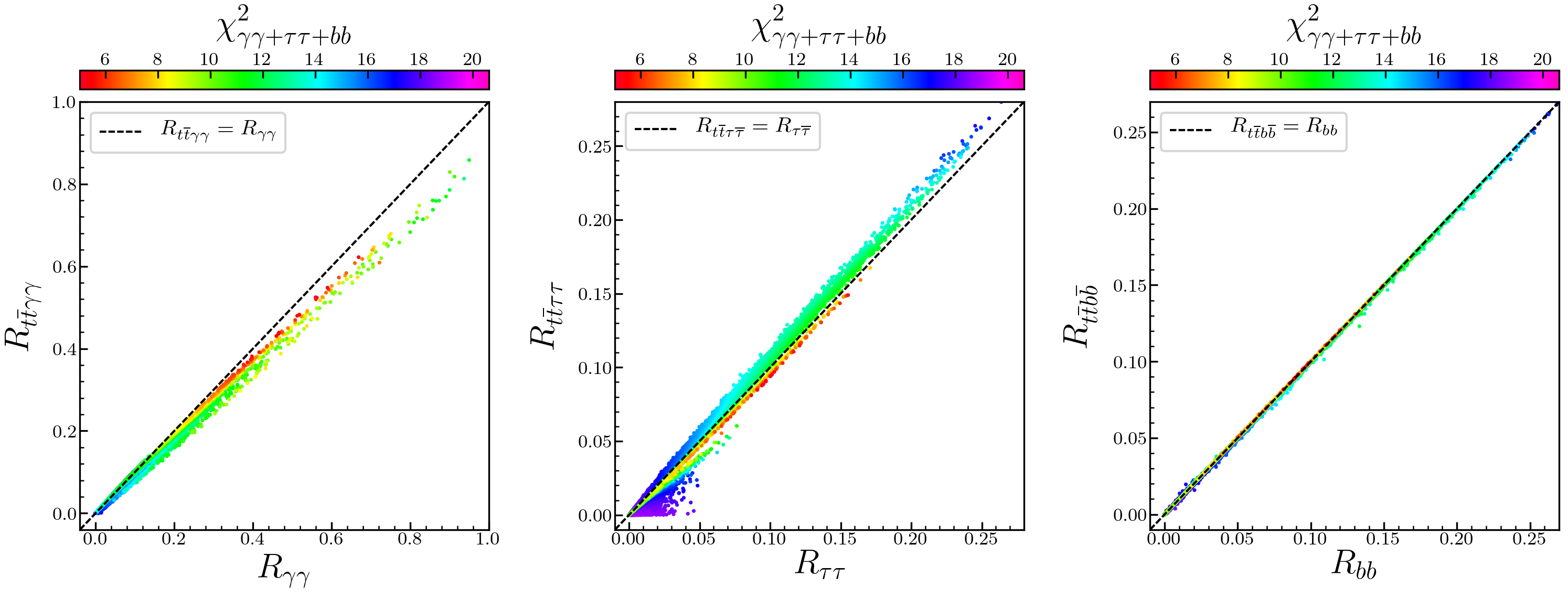}
\caption{Same as in Fig. \ref{fig1}, but on the planes of signal strengths in top-quark-pair associated channels versus these of existing excess channels: $R_{t\bar{t}\gamma\gamma}$ versus $R_{\gamma\gamma}$ (left), $R_{t\bar{t}\tau\tau}$ versus $R_{\tau\tau}$ (middle) and $R_{t\bar{t}b\bar{b}}$ versus $R_{bb}$ (right) planes.}
\label{fig4}
\end{figure*}

In Fig. \ref{fig1}, we project the surviving samples on the signal strengths $R_{\gamma \gamma}$($gg \!\to\! h_1 \!\to\! \gamma\gamma$), $R_{\tau \tau}$($gg \!\to\! h_1 \!\to\! \tau\bar{\tau}$) and $R_{bb}$($e^{+}e^{-} \!\to\! Zh_1 \!\to\! Zb\bar{b}$) versus width ratio $R_{\Gamma}$ (total decay width of $h_1$ divided by that of a SM Higgs of the same mass) planes, with colors denoting $\chi^2_{\gamma\gamma+\tau\tau+bb}$. 
From this figure, one can see that the low-mass excess data are powerful in distinguishing the surviving samples. 
For the $2\sigma$ samples, $R_{\Gamma}\lesssim 0.1$, $0.2\lesssim R_{\gamma\gamma}\lesssim 0.8$, and $R_{\tau\tau}, R_{b b} \lesssim0.2$. 
The surviving samples can be sorted into two regions obviously: the $R_{\Gamma}\lesssim 0.1$ region and the $R_{\gamma\gamma}\lesssim 0.2$ region. 
Note that the $2\sigma$ samples can only locate in the former. 
Hereafter to compare with the $2\sigma$ samples in the former region, we also consider the $3\sigma$ samples, or samples with $8.03\lesssim  \chi^2_{\gamma\gamma+\tau\tau+bb}\lesssim 14.16$, in the latter region, calling them  small-$R_{\gamma\gamma}$ samples. 
From the middle and right planes, one can see that for the $2\sigma$ samples $0.04\lesssim R_{\tau\tau}, R_{bb}\lesssim 0.16$, while for the small-$R_{\gamma\gamma}$ samples $0.05\lesssim R_{\tau\tau}, R_{bb}\lesssim 0.25$. 
Combining with the experimental data, one can know that the $2\sigma$ samples mainly fit well with the CMS di-photon excess, and small-$R_{\gamma\gamma}$ samples mainly fit well with the LEP $Zb\bar{b}$ excess. 
The CMS di-tau excess has so large uncertainty that it can not be dominant in our samples.

The signal strengths are related to the reduced couplings by 
\begin{eqnarray}
&R_{\gamma \gamma} = c_g^2c_{\gamma}^2/R_{\Gamma} \,, \nonumber\\ 
&R_{\tau\tau} = c_g^2c_{\tau}^2/R_{\Gamma} \,, \nonumber\\ 
&R_{bb} = c_V^2 c_b^2/R_{\Gamma} \,. 
\end{eqnarray}
So in Fig. \ref{fig2}, we project the surviving samples on the signal strengths versus reduced coupling planes, also with colors denoting $\chi^2_{\gamma\gamma+\tau\tau+bb}$. 
From this figure, one can see that the reduced couplings can be sorted into two classes: 
$|c_{\gamma}| \approx |c_g| \approx |c_t| \approx |c_V|$, and $c_b \approx c_{\tau}$. 
And it can also be found that the width ratio is determined by $c_b^2$, for the dominant branching ratio of the light scalar is that to $b\bar{b}$. 
Thus one can rewrite the signal strengths approximately to 
\begin{eqnarray}
&&R_{\gamma \gamma} \approx c_t^4 / c_b^2 \,,   \nonumber\\ 
&&R_{\tau\tau} \approx c_t^2 \nonumber\,, \\ 
&&R_{bb} \approx c_t^2 \,, 
\label{ratios}
\end{eqnarray}
where one can see that small width ratio $R_{\Gamma}$, or approximate $c_b^2$, can increase the di-photon rate, but can not increase the $b\bar{b}$ and di-tau rates.
Then $\chi^2_{\gamma\gamma+\tau\tau+bb}$ can be  approximately written to 
\begin{eqnarray}
\chi^2_{\gamma\gamma+\tau\tau+bb} \approx && \hspace{-6mm} \left[25\left(\frac{c_t}{c_b}\right)^4 + 311.8 \right] c_t^4 \nonumber \\
&& \hspace{-6mm} - \left[30\left(\frac{c_t}{c_b}\right)^2 + 81.6 \right] c_t^2 + 19.0 \,. ~~~~~~
\label{chi2sim}
\end{eqnarray}

From Fig. \ref{fig2} one can also see that: for $2\sigma$ samples, the light scalar has negative reduced couplings to fermions and W/Z bosons, with $0.3\lesssim -c_t \lesssim 0.4$ and $0.05 \lesssim -c_b \lesssim 0.3$;  
for small-$R_{\gamma\gamma}$ samples, the reduced couplings are positive, with $0.25\lesssim c_t \lesssim 0.45$ and $0.25 \lesssim c_b \lesssim 1$. 
One can also check Eq.(\ref{chi2sim}) simply: when $c_t=0$, or $c_t=0.5$ with $c_b=1$, $\chi^2_{\gamma\gamma+\tau\tau+bb} \approx 19$; 
when $|c_t|=2|c_b|=\sqrt{0.1}$, $\chi^2_{\gamma\gamma+\tau\tau+bb} \approx 6$.

In Fig. \ref{fig3}, we project the surviving samples on the $S_{12}$-$S_{11}$, $\tan\beta$-$S_{12}$ and $c_t$-$c_b$ planes. 
From Fig. \ref{fig3} one can see that when $c_t, c_b \lesssim 0$, or the couplings to quarks are flipped in sign, $|c_t/c_b| \gtrsim 1$, where most $2\sigma$ samples are located in; otherwise $|c_t/c_b| \lesssim 1$, then $R_{\gamma\gamma}$ will be smaller.
Combining Fig. \ref{fig3} and Eq. (\ref{h1coup}) one can see that, because $\tan\beta\gg 1$ and $|S_{12}|\ll 1$, we can safely have 
\begin{eqnarray}
&&c_V \approx c_t \approx S_{11} \,, \nonumber\\
&&c_b \approx S_{12}\tan\beta \,.
\end{eqnarray}
From Fig. \ref{fig3}, one can also see that for the $2\sigma$ samples, $|S_{11}| \gg |S_{12}|$, which means that the lightest Higgs boson is mainly mixed by the singlet and up-type doublet fields. Different from this, in the wrong sign limit \cite{Ferreira:2014naa, Ferreira:2014qda} in the type-II two Higgs doublet model, the lighter Higgs boson is mixed by the up- and down-type doublets fields.
And we also checked that the missing of $c_t \gtrsim c_b$ case in  Fig. \ref{fig3} is because we choose positive $\mu_{\rm eff}$, which is favored by the muon $g-2$ constraint. 
For the down-type doublet-like Higgs boson in NMSSM needs to be much heavier than the other two Higgs bosons to escape the constraints, $S_{12}$, or the mixing between singlet and down-type doublet, should be very small compared with $S_{11}$, thus the case of $c_t \lesssim 0$ but $c_b \gtrsim 0$ is also not very favored. 

Considering the mass region of excesses are close to the $Z$ boson mass, we consider the scalar's production associated with a top quark pair to reduce the backgrounds, with the signal strengths written as 
\begin{eqnarray}
&R_{t\overline t\gamma\gamma} = c_t^2c_{\gamma}^2/R_{\Gamma} \,, \nonumber\\
&R_{t\overline t\tau \overline\tau} = c_t^2c_{\tau}^2/R_{\Gamma} \,,\nonumber\\ 
&R_{t\overline tb\overline b} = c_t^2 c_b^2/R_{\Gamma} \,. 
\end{eqnarray}
In Fig. \ref{fig4}, we project the surviving samples on the planes of signal strengths of top-quark-pair associated channels versus these of three excess channels, respectively. 
From this figure one can see that $R_{t\bar{t}\gamma\gamma}\approx R_{\gamma\gamma}$, $R_{t\bar{t}\tau\bar{\tau}}\approx R_{\tau\tau}$, $R_{t\bar{t}b\bar{b}}\approx R_{bb}$. 
There is a small difference, especially between top-pair-associated and gluon-gluon-fusion channels. 
For $2\sigma$ samples, the latter is slightly larger than the former; while for the small-$R_{\gamma\gamma}$ samples, the former is slightly larger than the latter. 
The difference comes from the contributions of squarks, and they are positive or negative depending on $c_t$, the reduced couplings to the top quark. 
And the difference is small because of the high mass bounds of squarks \cite{Wang:2021lwi} from SUSY search results. 
As a comparison, new light colored particles can contribute much to the gluon-gluon-fusion channel \cite{Cao:2013wqa}. 

\begin{table*}[htbp]
	\centering
	\caption{Eight benchmark points for the surviving samples. }
	\label{tab2}
	\begin{tabular}{ccccccccc}
		\toprule
		\hline
		~~~~~~~~~~~~~~~~~~& ~~~~~~P1~~~~~~ & ~~~~~~P2~~~~~~ & ~~~~~~P3~~~~~~ & ~~~~~~P4~~~~~~ & ~~~~~~P5~~~~~~ & ~~~~~~P6~~~~~~ & ~~~~~~P7~~~~~~ & ~~~~~~P8~~~~~~ \\
		\hline
		$\lambda$ &0.315&0.348&0.335&0.271&0.297&0.165&0.116&0.339\\
		$\kappa$ &0.128&0.138&0.102&0.052&-0.121&-0.051&0.044&0.544 \\
		tan$\beta$ &30.8&30.5&31.8&25.1&21.2&6.4&11.1&14.7 \\
		$\mu_{\rm eff}$/GeV &272&284&308&263&288&391&214&232\\
		$M_0$/GeV &1503&1824&2342&1954&3415&303&790&335 \\
		$A_0$/GeV &1804&1947&867&1081&949&-1717&1653&1617\\
		$M_1$/GeV &-49.6&-50.1&-19.7&-18.1&-83.9&-75.4&-63.7&-622 \\
		$M_2$/GeV &-2061&-2363&-3892&-2862&-302&249&-84&742 \\
		$M_3$/GeV &2877&3037&4948&4920&2992&1439&2004&2574\\
		$A_\lambda$/GeV &7881&8510&8077&4474&6395&2212&2807&2745\\
		$A_\kappa$/GeV &1610&2224&2111&837&1797&572&-107&-3538\\
		$m_{h_1}$/GeV  &96.5&95.0&95.2&95.6&98.3&96.9&98.8&96.4\\
		$m_{h_2}$/GeV  &124.9&125.2&126.0&125.7&125.9&125.7&126.1&126.0\\
		$S_{11}$ &-0.343&-0.340&-0.287&-0.255&-0.220&0.383&0.399&0.336\\
		$S_{12}$ &-0.0050&-0.0046&-0.0033&-0.0033&-0.0023&0.0695&0.0438&0.0405\\
		$c_t$ &-0.344&-0.341&-0.287&-0.256&-0.220&0.388&0.400&0.337\\
		$c_V$ &-0.343&-0.340&-0.287&-0.255&-0.220&0.390&0.401&0.338\\
		$c_b$ &-0.153&-0.139&-0.104&-0.082&-0.050&0.451&0.489&0.597\\
		$c_\tau$ &-0.153&-0.139&-0.104&-0.082&-0.050&0.451&0.489&0.597\\
		$|c_g|$ &0.356&0.354&0.299&0.268&0.232&0.385&0.395&0.324\\
		$|c_\gamma|$ &0.382&0.381&0.324&0.291&0.255&0.377&0.376&0.288\\
		$R_\Gamma$ &0.0200&0.0172&0.0107&0.0074&0.0046&0.1178&0.1396&0.1953\\
		$R_{\gamma\gamma}$  &0.548&0.618&0.510&0.472&0.455&0.106&0.096&0.026\\
		$R_{\tau\tau}$ &0.088&0.082&0.053&0.037&0.017&0.152&0.162&0.113\\
		$R_{bb}$ &0.083&0.077&0.049&0.035&0.016&0.156&0.167&0.123\\
		$R_{t\bar{t}\gamma\gamma}$  &0.509&0.571&0.469&0.431&0.410&0.108&0.098&0.029\\
		$R_{t\bar{t}\tau\bar{\tau}}$  &0.082&0.076&0.049&0.034&0.016&0.155&0.166&0.123\\
		$R_{t\bar{t}b\bar{b}}$  &0.083&0.077&0.050&0.035&0.016&0.154&0.166&0.122\\
		$\chi^2_{\gamma\gamma+\tau\tau+bb}$   &5.37&5.50&6.87&7.91&9.26&10.95&11.42&12.96\\
		$\chi_{125}^2$ &116.4&116.7&103.8&99.1&95.7&99.2&99.4&90.9\\
		$P_{125}$ &0.344&0.337&0.673&0.784&0.850&0.782&0.777&0.919\\
		$m_{\tilde{\chi}^0_1}$ &44.53&46.01&43.15&43.00&60.38&43.92&44.98&227.96\\
		$\Omega h^2$ &0.0213&0.0719&0.0566&0.0187&0.0524&0.0862&0.0086&0.0065\\
		$Br(h_2 \to \tilde{\chi}^0_1 \tilde{\chi}^0_1)$ &0.0027\%&0.0016\%&0.022\%&0.085\%&0.15\%&1.03\%&0.028\%&0.00\%\\
		\hline
		\bottomrule
	\end{tabular}
\end{table*}

In Tab. \ref{tab2}, we list the detailed information of eight representative benchmark points for further study, where $\chi^2_{125}$ and $P_{125}$ are the chi-square and P value from 125 GeV Higgs data of 111 groups (the number of degrees of freedom is 111). Note that for a SM Higgs of $125.09\GeV$, $\chi^2_{125}=89.7$ and $P_{125}=0.932$.
From this table, one can see that it is difficult to satisfy the 125 GeV Higgs data and the $95\sim 100\GeV$ excesses simultaneously at $2\sigma$ level. 
E.g., for Point P4, with the $95\sim 100\GeV$ excesses globally satisfied at $2\sigma$ level the 125 GeV Higgs data can be at $78.4\%$, worse than that of a SM Higgs boson of $125\GeV$. 

In the end, we add discussions on dark matter, invisible Higgs decay, and electroweakino searches. 
\begin{itemize}
\item For benchmark points P1-P7, the dark matter is bino-like, and the main annihilation mechanism is $Z/h_2$ funnel. 
The mass of dark matter is different from $M_1$, because the parameters $M_{1,2,3}$ are defined at the GUT scale. 
There are correlations between parameters at GUT and SUSY scales, similar to those we present in Appendix A of our former work \cite{Wang:2018vrr}. 
\item We have considered the constraint of invisible Higgs decay with the code {\sf HiggsBounds}, where the corresponding experimental data in Refs. \cite{ATLAS:2019nkf, CMS:2018uag} are included. 
For benchmark points P1-P7, the invisible Higgs decay $Br(h_2\to\tilde{\chi}^0_1 \tilde{\chi}^0_1)$ are all smaller than about $1\%$, because the large invisible ratios are not favored by both $125$ and $95\sim100\GeV$ Higgs data. 
\item We have also imposed constraints from SUSY searches with the code {\sf SModelS}. 
For benchmark point P8, the dark matter is Higgsino-like and the main annihilation mechanism is $W^\pm/Z$ exchanges. 
This point can escape the constraints from searches for electroweakinos in Ref. \cite{ATLAS:2021yqv} because of its compressed mass spectrum and multiple decay modes. 
In the low mass region, it has SUSY particles like Higgsino-like charginos and neutralinos of about $230\GeV$, bino-like neutralino of $390\GeV$, wino-like charginos and neutralinos of about $590\GeV$, $\tilde{\tau}_1$ of $246\GeV$, $\tilde{\nu}_{\tau}$ of $353\GeV$, $\tilde{\mu}_1$ of $478\GeV$, $\tilde{\nu}_{\mu}$ of $472\GeV$, etc. 
\end{itemize}

\section{Conclusions}
\label{sec:conc}
In this work, we consider a light Higgs boson in the NMSSM to interpret the CMS di-photon and di-tau excesses, and LEP $b\bar{b}$ excess, in the $95\sim100\GeV$ mass region. 
We first scan the parameter space and consider a series of constraints, including these of Higgs, dark matter, SUSY searches, etc. 
Then for each surviving sample, we calculate a chi-square considering its global fit to the three excess data. 
And we focus on two respective kinds of samples: the $2\sigma$ samples and small-$R_{\gamma\gamma}$ samples. 
Finally, we get the following conclusions: 
\begin{itemize}
\item In NMSSM it is difficult to satisfy the $95\sim 100\GeV$ excesses simultaneously (not possible globally at $1\sigma$ level, or simultaneously at $2\sigma$ level).
\item The light Higgs boson's global fit to the three excesses is mainly determined by its couplings to up- and down-type fermions, which can be approximately written as in Eq. (\ref{chi2sim}). 
\item The globally $2\sigma$ samples have negative reduced couplings to fermions and massive vector bosons, while they are positive for the small-$R_{\gamma\gamma}$ ones.
\item The globally $2\sigma$ samples have decay width smaller than one-tenth of the corresponding SM value, which can increase its di-photon rate but can not increase its di-tau rate.  
\item The small-$R_{\gamma\gamma}$ samples can have $Zb\bar{b}$ signal right fit to the LEP $b\bar{b}$ excess, but have smaller di-photon and di-tau rates.
\item The top-quark-pair associated signal strengths are nearly equal to these of the three exciting excesses, respectively. 
\end{itemize}

\section*{Acknowledgments.}
\textit{Acknowledgements}. This work was supported by the National Natural Science Foundation of China (NNSFC) under Grant Nos. 12275066, 11605123, 11547103, and 12074295.

\section*{Appendix A: Error level of Eq. (\ref{chi2sim})}

\begin{figure}[H]
	\centering
	\includegraphics[width=0.46\textwidth]{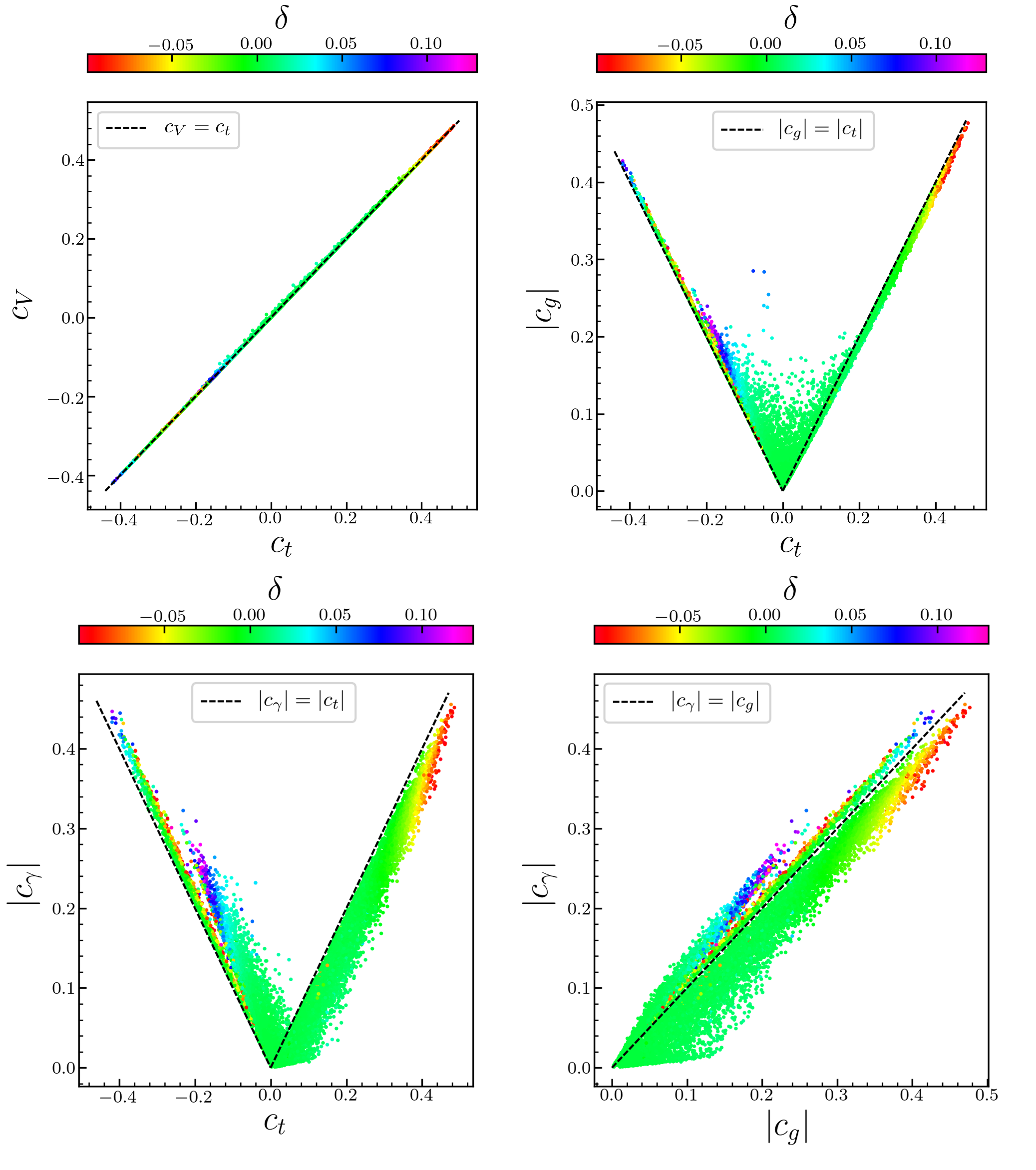}
	\caption{Surviving samples on the planes of $c_V$ (upper left), $|c_g|$ (upper right) and $|c_{\gamma}|$ (lower left) versus $c_t$, and $|c_\gamma|$ versus $|c_g|$ (lower right), respectively, with colors indicating the error ratio $\delta$ between the approximated and complete ones in Eq. (\ref{chi2sim}). }
	\label{figa} 
\end{figure}

To show the error level between the approximated chi-square and complete ones in Eq. (\ref{chi2sim}), we give Fig. \ref{figa}. 
In this figure, one can see clearly that $c_V\approx c_t$ is a very good approximation. 
For most samples, since the charged Higgs bosons and most SUSY particles are heavy, their contributions to the loop-induced couplings $c_g$ ($c_\gamma$) are much smaller than these of the SM particles top quark (and $W$ boson). 
Thus $c_g$ ($c_\gamma$) are mainly determined by the top quark's coupling $c_t$ (and the $W$ boson's coupling $c_V$). 
And since $c_V\approx c_t$, for most samples we have $c_\gamma \approx c_V\approx c_t \approx c_g$. 
From Fig. \ref{figa}, one can also see that, the error level between the approximate chi-square and complete ones in Eq. (\ref{chi2sim}) are below $5\%$ for most samples, and at most about $15\%$ for all ones. 
Therefore, Eqs. (\ref{chi2sim}) and (\ref{ratios}) are good approximations for most samples, and are interesting with only two variable quantities.

\end{multicols}

\begin{center}
\rule[10pt]{10cm}{0.2mm}
\end{center}
\begin{multicols}{2}

\end{multicols}

\end{document}